\documentclass{article}%
\usepackage{amsfonts}
\usepackage{amsmath}
\usepackage{amssymb}
\usepackage{charter}
\usepackage{graphicx}
\usepackage{cite}%
\providecommand{\U}[1]{\protect \rule{.1in}{.1in}}

\begin{document}

\title{Atomic coherent state in Schwinger bosonic realization for optical Raman
coherent effect}
\author{Hong-yi Fan$^{1},$Xue-xiang Xu$^{1,2}$, and Li-yun Hu$^{2}$%
\thanks{{\small Corresponding author. E-mail address: hlyun2008@126.com.}}\\$^{1}${\small Department of Physics, Shanghai Jiao Tong University, Shanghai,
200030, China;}\\$^{2}${\small College of Physics and Communication Electronics, Jiangxi Normal
University, Nanchang, 330022, China.}}
\maketitle

\begin{abstract}
For optical Raman coherent effect we introduce the atomic coherent state (or
the angular momentum coherent state with various angular momemtum values) in
Schwinger bosonic realization, they are the eigenvectors of the Hamiltonian
describing the Raman effect. Similar to the fact that the photon coherent
state describes laser light, the atomic coherent state is related to Raman process.

\end{abstract}

\section{Introduction}

Atomic coherent states (or the angular momentum coherent state with various
angular momemtum values) are sometimes referred to in the literature as spin
coherent states or Bloch states \cite{1,2,3,4,5,6}. They have been
successfully applied to many branches of physics \cite{7,8,9,10}. For example,
Arecchi et al. applied atomic coherent states to describe interactions between
radiation field and an assembly of two-level atoms \cite{4}. Narducci, Bowden,
Bluemel, Garrazana and Tuft \cite{7} used atomic coherent state to study
multitime correlation function for systems with observables satisfying an
angular momentum algebra, which suggested a convenient classical-quantum
correspondence rule for angular momentum degrees of freedom. Takahashi and
Shibata \cite{9} transformed some equation of motion for density matrix of a
damped spin system into that of a quasi-distribution. Gerry and Benmoussa
\cite{10} have studied the generation of spin squeezing by the repeated action
of the angular momentum Dicke lowering operator on an atomic coherent state.
In this work we shall introduce the atomic coherent state in Schwinger bosonic
realization to study Raman coherent effect in the context of quantum optics.

It is known that the Raman coherent effect, a monochromatic light wave
incident on a Raman active medium gives rise to a parametric coupling between
an optical vibrational mode and the mode of the radiation field, the so-called
Stocks mode. (In the case of Brillouin scattering, there is a similar
coupling, where the vibrations are at acoustical, rather than optical
frequencies.) The simplest Hamiltonian model for describing Raman coherent
effect is%

\begin{equation}
H=\omega_{1}a^{\dagger}a+\omega_{2}b^{\dagger}b-i\lambda \left(  a^{\dagger
}b-ab^{\dagger}\right)  , \label{1}%
\end{equation}
which is a two coupled oscillator model. In this work we shall show that the
atomic coherent state (some assembly of angular momentum states, so named
angular momentum coherent state) expressed in terms of Schwinger bosonic
realization of angular momentum \cite{11} has its obvious physical background,
i.e., a set of energy eigenstates of two coupled bosonic oscillators with the
Hamiltonian can be classified as the atomic coherent state $\left \vert
\tau \right \rangle _{j}$ according to the angular momentum value of $j$, where
$\tau$ is determined by the dynamic parameters $\omega_{1},\omega_{2},\lambda
$. Thus the Raman coherent effect is closely related to atomic coherent state
theory, while the laser is described by the coherent state theoretically.

\section{Brief review of the atomic coherent state (ACS) in Schwinger bosonic
realization}

The atomic coherent state with angular momentum value $j$ is defined as
\cite{4,5,6,7}%

\begin{equation}
\left \vert \tau \right \rangle =\exp(\mu J_{+}-\mu^{\ast}J_{-})\left \vert
j,-j\right \rangle =(1+\left \vert \tau \right \vert ^{2})^{-j}e^{\tau J_{+}%
}\left \vert j,-j\right \rangle , \label{2}%
\end{equation}
where $J_{+}$ is the raising operator of the angular momentum state
$\left \vert j,m\right \rangle $, $\left \vert j,-j\right \rangle $ is the lowest
weight state annihilated by $J_{-}$, and%

\begin{equation}
\mu=\frac{\theta}{2}\text{e}^{-\text{i}\varphi},\text{ }\tau=\text{e}%
^{-\text{i}\varphi}\tan(\frac{\theta}{2}). \label{3}%
\end{equation}
In the $j$-subspace the completeness relation for $\left \vert \tau
\right \rangle $ is%

\begin{equation}
\int \frac{\text{d}\Omega}{4\pi}\left \vert \tau \right \rangle \left \langle
\tau \right \vert =\sum_{m=-j}^{j}\left \vert j,m\right \rangle \left \langle
j,m\right \vert =1_{j}, \label{4}%
\end{equation}
where d$\Omega=\sin \theta$d$\theta$d$\varphi$, and%

\begin{equation}
\left \langle \tau^{\prime}\right.  \left \vert \tau \right \rangle =\frac
{(1+\tau^{\prime}\tau^{\ast})^{2j}}{(1+\left \vert \tau \right \vert ^{2}%
)^{j}(1+\left \vert \tau^{\prime}\right \vert ^{2})^{j}}. \label{5}%
\end{equation}
Using $[J_{+},J_{-}]=2J_{z},$ $[J_{\pm},J_{z}]=\mp J_{\pm}$, one can show that
$\left \vert \tau \right \rangle $ obeys the following eigenvector equations,%

\begin{align}
(J_{-}+\tau^{2}J_{+})\left \vert \tau \right \rangle  &  =2j\tau \left \vert
\tau \right \rangle ,\nonumber \\
(J_{-}+\tau J_{z})\left \vert \tau \right \rangle  &  =j\tau \left \vert
\tau \right \rangle ,\label{6}\\
(\tau J_{+}-J_{z})\left \vert \tau \right \rangle  &  =j\left \vert \tau
\right \rangle .\nonumber
\end{align}
Employing the Schwinger Bose operator realization of angular momentum%

\begin{equation}
J_{+}=a^{\dagger}b,\text{ }J_{-}=ab^{\dagger},\text{ }J_{z}=\frac{1}{2}\left(
a^{\dagger}a-b^{\dagger}b\right)  , \label{7}%
\end{equation}
where $[a,a^{\dagger}]=1,$ $[b,b^{\dagger}]=1$ and $\left \vert
j,m\right \rangle \ $is realized as%
\begin{align}
\left \vert j,m\right \rangle  &  =\frac{a^{\dagger j+m}b^{\dagger j-m}}%
{\sqrt{(j+m)!(j-m)!}}\left \vert 00\right \rangle \nonumber \\
&  =\left \vert j+m\right \rangle \otimes \left \vert j-m\right \rangle ,\text{
\  \ }(m=-j,\cdots,j), \label{8}%
\end{align}
note that the last ket is written in two-mode Fock space, then $\left \vert
j,-j\right \rangle =\left \vert 0\right \rangle \otimes \left \vert 2j\right \rangle
,$ and the atomic coherent state $\left \vert \tau \right \rangle $ is expressed as%

\begin{align}
\left \vert \tau \right \rangle  &  =e^{\mu J_{+}-\mu^{\ast}J_{-}}\left \vert
0\right \rangle \otimes \left \vert 2j\right \rangle \nonumber \\
&  =\frac{1}{\sqrt{(2j)!}}[b^{\dagger}\cos(\frac{\theta}{2})+a^{\dagger
}\text{e}^{-\text{\texttt{i}}\varphi}\sin(\frac{\theta}{2})]^{2j}\left \vert
00\right \rangle \nonumber \\
&  =\frac{1}{\left(  1+\left \vert \tau \right \vert ^{2}\right)  ^{j}}\sum
_{l=0}^{2j}\sqrt{\frac{(2j)!}{l!(2j-l)!}}\tau^{2j-l}\left \vert
2j-l\right \rangle \otimes \left \vert l\right \rangle \label{9}%
\end{align}
Especially when $j=0$, $\left \vert \tau \right \rangle =\left \vert
00\right \rangle $ is just the two-mode vacuum state in Fock space. Using the
normal ordering form of the two-mode vacuum projector $\left \vert
00\right \rangle \left \langle 00\right \vert =:e^{-a^{\dagger}a-b^{\dagger}b}:$,
we can use the technique of integration within an ordered product of operators
\cite{12,13} to prove in $j$-subspace,
\begin{align}
\int \frac{\text{d}\Omega}{4\pi}\left \vert \tau \right \rangle \left \langle
\tau \right \vert  &  =\frac{1}{\left(  2j\right)  !}\int_{0}^{\pi}d\theta
\sin \theta \int_{0}^{2\pi}d\phi:\left(  b^{\dagger}\cos \frac{\theta}%
{2}+a^{\dagger}e^{-i\phi}\sin \frac{\theta}{2}\right)  ^{2j}\nonumber \\
&  \times \left.  \left(  b\cos \frac{\theta}{2}+ae^{i\phi}\sin \frac{\theta}%
{2}\right)  ^{2j}\exp \left(  -a^{\dagger}a-b^{\dagger}b\right)  :\right.
\nonumber \\
&  =:\frac{\left(  a^{\dagger}a+b^{\dagger}b\right)  ^{2j}}{\left(
2j+1\right)  !}e^{-a^{\dagger}a-b^{\dagger}b}:, \label{10}%
\end{align}
the completeness relation of $\left \vert \tau \right \rangle $ in the whole
two-mode Fock space can be obtained after summing over $j$:%
\begin{align}
&  \sum_{2j=0}^{\infty}(2j+1)\int \frac{\text{d}\Omega}{4\pi}\left \vert
\tau \right \rangle \left \langle \tau \right \vert \nonumber \\
&  =\sum_{2j=0}^{\infty}:\frac{\left(  a^{\dagger}a+b^{\dagger}b\right)
^{2j}}{\left(  2j\right)  !}e^{-a^{\dagger}a-b^{\dagger}b}:=1, \label{11}%
\end{align}
which means that atomic coherent states in Schwinger bosonic realization with
all values of $j$ forms a complete set.

\section{Atomic coherent state as energy eigenstates of H}

Now we inquire whether the atomic coherent state with a definite angular
momentum value $j$ is the solution of the stationary Schrodinger equation%
\begin{equation}
H\left \vert \tau \right \rangle =E\left \vert \tau \right \rangle . \label{12}%
\end{equation}
In order to solve Eq.(\ref{12}) we directly use Eq.(\ref{9}) and the relation%
\begin{equation}
a^{\dagger}\left \vert n\right \rangle =\sqrt{n+1}\left \vert n+1\right \rangle
,\text{ }a\left \vert n\right \rangle =\sqrt{n}\left \vert n-1\right \rangle ,
\label{13}%
\end{equation}
to calculate%
\begin{align}
H\left \vert \tau \right \rangle  &  =\frac{1}{\left(  1+\left \vert
\tau \right \vert ^{2}\right)  ^{j}}\sum_{l=0}^{2j}\sqrt{\frac{(2j)!}%
{l!(2j-l)!}}\left[  \omega_{1}\left(  2j-l\right)  +\omega_{2}l\right]
\tau^{2j-l}\left \vert 2j-l\right \rangle \otimes \left \vert l\right \rangle
\nonumber \\
&  -i\lambda \frac{1}{\left(  1+\left \vert \tau \right \vert ^{2}\right)  ^{j}%
}\sum_{l=1}^{2j}\sqrt{\frac{(2j)!}{\left(  l-1\right)  !(2j-l+1)!}}\left(
2j-l+1\right)  \tau^{2j-l}\left \vert 2j-l+1\right \rangle \otimes \left \vert
l-1\right \rangle \nonumber \\
&  +i\lambda \frac{1}{\left(  1+\left \vert \tau \right \vert ^{2}\right)  ^{j}%
}\sum_{l=0}^{2j-1}\sqrt{\frac{(2j)!}{\left(  l+1\right)  !(2j-l-1)!}}%
\tau^{2j-l}\left(  l+1\right)  \left \vert 2j-l-1\right \rangle \otimes
\left \vert l+1\right \rangle . \label{14}%
\end{align}
Let $l\mp1\rightarrow l$ in the second and third term of the r.h.s. of
Eq.(\ref{14}), respectively, we have%
\begin{align}
H\left \vert \tau \right \rangle  &  =\frac{1}{\left(  1+\left \vert
\tau \right \vert ^{2}\right)  ^{j}}\sum_{l=0}^{2j}\sqrt{\frac{(2j)!}%
{l!(2j-l)!}}\tau^{2j-l}\left \{  \left[  \omega_{1}\left(  2j-l\right)
+\omega_{2}l\right]  -i\lambda \left(  2j-l\right)  \frac{1}{\tau}+i\lambda \tau
l\right \}  \left \vert 2j-l\right \rangle \otimes \left \vert l\right \rangle
\nonumber \\
&  =\frac{1}{\left(  1+\left \vert \tau \right \vert ^{2}\right)  ^{j}}\sum
_{l=0}^{2j}\sqrt{\frac{(2j)!}{l!(2j-l)!}}\tau^{2j-l}\left \{  2\left(
\omega_{1}-i\frac{\lambda}{\tau}\right)  j+\left[  \left(  \omega_{2}%
-\omega_{1}\right)  +i\lambda \left(  \tau+\frac{1}{\tau}\right)  \right]
l\right \}  \left \vert 2j-l\right \rangle \otimes \left \vert l\right \rangle
\nonumber \\
&  =2\left(  \omega_{1}-i\frac{\lambda}{\tau}\right)  j\left \vert
\tau \right \rangle +\frac{1}{\left(  1+\left \vert \tau \right \vert ^{2}\right)
^{j}}\sum_{l=0}^{2j}\sqrt{\frac{(2j)!}{l!(2j-l)!}}\tau^{2j-l}\left[  \left(
\omega_{2}-\omega_{1}\right)  +i\lambda \left(  \tau+\frac{1}{\tau}\right)
\right]  l\left \vert 2j-l\right \rangle \otimes \left \vert l\right \rangle .
\label{15}%
\end{align}
We see when the following condition is satisfied,%
\begin{equation}
i\lambda \tau^{2}+\tau \left(  \omega_{2}-\omega_{1}\right)  +i\lambda
=0\Rightarrow \tau_{\pm}=\frac{\left(  \omega_{1}-\omega_{2}\right)  \pm
\sqrt{\left(  \omega_{1}-\omega_{2}\right)  ^{2}+4\lambda^{2}}}{2i\lambda}.
\label{16}%
\end{equation}
then $\left \vert \tau_{\pm}\right \rangle ,$ expressed by Eq.(\ref{9}), is the
eigenstate of $H$ with eigenvalue%
\begin{align}
E  &  =2\left(  \omega_{1}-i\frac{\lambda}{\tau}\right)  j\nonumber \\
&  =j\left[  \left(  \omega_{1}+\omega_{2}\right)  \pm \sqrt{\left(  \omega
_{1}-\omega_{2}\right)  ^{2}+4\lambda^{2}}\right]  \label{17}%
\end{align}
Hence $H$'s eigenvectors are classifiable according to the angular momentum
value $j$. Especially, when $\omega_{1}=\omega_{2}=\omega$, from
Eqs.(\ref{16})-(\ref{17}) we know $\tau_{\pm}=\mp i,$ $E_{\pm}=2j\left(
\omega \pm \lambda \right)  .$

\section{Some fundamental atomic coherent states as H's eigenstates}

We now investigate some fundamental atomic coherent states as $H$'s
eigenstates. In the case of $j=1/2,$ from Eq.(\ref{9}) we know the eigenstate
of $H$ is
\begin{align}
\left \vert \tau_{\pm}\right \rangle _{j=1/2}  &  =\frac{1}{\left(  1+\left \vert
\tau_{\pm}\right \vert ^{2}\right)  ^{1/2}}\left(  \tau_{\pm}\left \vert
1\right \rangle \otimes \left \vert 0\right \rangle +\left \vert 0\right \rangle
\otimes \left \vert 1\right \rangle \right) \nonumber \\
\overset{\omega_{1}=\omega_{2}}{\rightarrow}\left \vert i_{\pm}\right \rangle
_{j=1/2}  &  =\frac{1}{\sqrt{2}}\left(  \mp i\left \vert 1\right \rangle
\otimes \left \vert 0\right \rangle +\left \vert 0\right \rangle \otimes \left \vert
1\right \rangle \right)  . \label{18}%
\end{align}
Indeed, one can check $H\left \vert i_{+}\right \rangle _{j=1/2}=\frac
{\omega+\lambda}{\sqrt{2}}\left(  -i\left \vert 1,0\right \rangle +\left \vert
0,1\right \rangle \right)  .$ In the case of $j=1,$%
\begin{align}
\left \vert \tau_{\pm}\right \rangle _{j=1}  &  =\frac{1}{1+\left \vert \tau
_{\pm}\right \vert ^{2}}\sum_{l=0}^{2}\sqrt{\frac{(2)!}{l!(2-l)!}}\tau_{\pm
}^{2-l}\left \vert 2-l\right \rangle \otimes \left \vert l\right \rangle
\nonumber \\
&  =\frac{1}{1+\left \vert \tau_{\pm}\right \vert ^{2}}\left(  \tau_{\pm}%
^{2}\left \vert 2\right \rangle \otimes \left \vert 0\right \rangle +\sqrt{2}%
\tau_{\pm}\left \vert 1\right \rangle \otimes \left \vert 1\right \rangle
+\left \vert 0\right \rangle \otimes \left \vert 2\right \rangle \right)
\nonumber \\
\overset{\omega_{1}=\omega_{2}}{\rightarrow}\left \vert i_{\pm}\right \rangle
_{1}  &  =\frac{1}{2}\left(  -\left \vert 2\right \rangle \otimes \left \vert
0\right \rangle \mp i\sqrt{2}\left \vert 1\right \rangle \otimes \left \vert
1\right \rangle +\left \vert 0\right \rangle \otimes \left \vert 2\right \rangle
\right)  . \label{19}%
\end{align}
In the case of $j=3/2,$%
\begin{align}
\left \vert \tau_{\pm}\right \rangle _{j=3/2}  &  =\frac{1}{\left(  1+\left \vert
\tau \right \vert ^{2}\right)  ^{3/2}}\left(  \tau_{\pm}^{3}\left \vert
3\right \rangle \otimes \left \vert 0\right \rangle +\sqrt{3}\tau_{\pm}%
^{2}\left \vert 2\right \rangle \otimes \left \vert 1\right \rangle +\sqrt{3}%
\tau_{\pm}\left \vert 1\right \rangle \otimes \left \vert 2\right \rangle
+\left \vert 0\right \rangle \otimes \left \vert 3\right \rangle \right)
\nonumber \\
\overset{\omega_{1}=\omega_{2}}{\rightarrow}\left \vert i_{\pm}\right \rangle
_{j=3/2}  &  =\frac{1}{2^{3/2}}\left(  \pm i\left \vert 3\right \rangle
\otimes \left \vert 0\right \rangle -\sqrt{3}\left \vert 2\right \rangle
\otimes \left \vert 1\right \rangle \mp i\sqrt{3}\left \vert 1\right \rangle
\otimes \left \vert 2\right \rangle +\left \vert 0\right \rangle \otimes \left \vert
3\right \rangle \right)  . \label{20}%
\end{align}
In the case of $j=2$%
\begin{align}
\left \vert \tau_{\pm}\right \rangle _{j=2}  &  =\frac{1}{\left(  1+\left \vert
\tau \right \vert ^{2}\right)  ^{2}}\sum_{l=0}^{4}\sqrt{\frac{4!}{l!(4-l)!}}%
\tau_{\pm}^{4-l}\left \vert 4-l\right \rangle \otimes \left \vert l\right \rangle
\nonumber \\
&  =\frac{1}{\left(  1+\left \vert \tau \right \vert ^{2}\right)  ^{2}}\left(
\tau_{\pm}^{4}\left \vert 4\right \rangle \otimes \left \vert 0\right \rangle
+2\tau_{\pm}^{3}\left \vert 3\right \rangle \otimes \left \vert 1\right \rangle
+\sqrt{6}\tau_{\pm}^{2}\left \vert 2\right \rangle \otimes \left \vert
2\right \rangle +2\tau_{\pm}\left \vert 1\right \rangle \otimes \left \vert
3\right \rangle +\left \vert 0\right \rangle \otimes \left \vert 4\right \rangle
\right) \nonumber \\
\overset{\omega_{1}=\omega_{2}}{\rightarrow}\left \vert i_{\pm}\right \rangle
_{j=2}  &  =\frac{1}{4}\left(  \left \vert 4\right \rangle \otimes \left \vert
0\right \rangle \pm2i\left \vert 3\right \rangle \otimes \left \vert 1\right \rangle
-\sqrt{6}\left \vert 2\right \rangle \otimes \left \vert 2\right \rangle
\mp2i\left \vert 1\right \rangle \otimes \left \vert 3\right \rangle +\left \vert
0\right \rangle \otimes \left \vert 4\right \rangle \right)  . \label{21}%
\end{align}
Thus we know how the eigenstate of $H$ is composed of the Fock states.

\section{Partition function and the Internal energy for H}

Knowing that $H$ is diagonal in the basis of atomic coherent state $\left \vert
\tau_{\pm}\right \rangle $, we can directly calculate its partition function by
virtue of its energy level.
\begin{align}
Z_{+}\left(  \beta \right)   &  =\mathtt{Tr}_{+}\left(  e^{-\beta H}\right)
=\sum_{2j=0}^{\infty}\left.  _{j}\left \langle \tau_{+}\right \vert e^{-\beta
H}\left \vert \tau_{+}\right \rangle _{j}\right. \nonumber \\
&  =\sum_{2j=0}^{\infty}e^{-\beta A2j}=\frac{1}{e^{\eta}-1}|_{\eta=-\beta
A}\nonumber \\
&  =\frac{1}{e^{-\beta A}-1}, \label{22}%
\end{align}
and
\begin{align}
Z_{-}\left(  \beta \right)   &  =\mathtt{Tr}_{-}\left(  e^{-\beta H}\right)
=\sum_{2j=0}^{\infty}\left.  _{j}\left \langle \tau_{-}\right \vert e^{-\beta
H}\left \vert \tau_{-}\right \rangle _{j}\right. \nonumber \\
&  =\frac{1}{e^{-\beta B}-1} \label{23}%
\end{align}
where
\begin{align}
A  &  =\frac{\left(  \omega_{1}+\omega_{2}\right)  +\sqrt{\left(  \omega
_{1}-\omega_{2}\right)  ^{2}+4\lambda^{2}}}{2},\nonumber \\
B  &  =\frac{\left(  \omega_{1}+\omega_{2}\right)  -\sqrt{\left(  \omega
_{1}-\omega_{2}\right)  ^{2}+4\lambda^{2}}}{2}. \label{24}%
\end{align}
satisfying $H\left \vert \tau_{+}\right \rangle =2Aj\left \vert \tau
_{+}\right \rangle ,H\left \vert \tau_{-}\right \rangle =2Bj\left \vert \tau
_{-}\right \rangle .$ Thus the total partition function is%
\begin{equation}
Z\left(  \beta \right)  =Z_{+}\left(  \beta \right)  Z_{-}\left(  \beta \right)
=\left(  \frac{1}{e^{-\beta A}-1}\right)  \left(  \frac{1}{e^{-\beta B}%
-1}\right)  , \label{25}%
\end{equation}
\bigskip and the internal energy of system is%
\begin{align}
\left \langle H\right \rangle _{e}  &  =-\frac{\partial}{\partial \beta}\ln
Z\left(  \beta \right) \nonumber \\
&  =-\frac{\partial}{\partial \beta}\left[  \ln \left(  \frac{1}{e^{-\beta A}%
-1}\right)  +\ln \left(  \frac{1}{e^{-\beta B}-1}\right)  \right] \nonumber \\
&  =\frac{A}{e^{A\beta}-1}+\frac{B}{e^{\beta B}-1}. \label{26}%
\end{align}

In summary, similar to the fact that the photon coherent state describes laser
light, the atomic coherent state is useful to classify the energy eigenstates
of the Hamiltonian describing the Raman effect.This may be useful to further
study stimulated Raman scattering since the scattered light behaves as laser light.

\textbf{ACKNOWLEDGEMENT:} We sincerely thank the referees for their
constructive suggestion. Work supported by the National Natural Science
Foundation of China under grants: 10775097 and 10874174, and the Research
Foundation of the Education Department of Jiangxi Province of China.

\end{document}